\documentclass[prb,aps,twocolumn,amsmath,amssymb,floatfix,superscriptaddress]{revtex4}

\usepackage{graphicx} 
\usepackage{float}
\usepackage{braket}
\usepackage{color}
\definecolor{dred}{rgb}{0.75,0,0}
\usepackage{soul}
\usepackage{multirow}
\usepackage[colorlinks=true, citecolor=blue, urlcolor=blue ]{hyperref}
\begin{document}

\title{Interaction-controlled localization in one-dimensional chain: From edges to domain walls}

\author{Rahul Samanta}
\email{rahul99$_$r@isical.ac.in}
\affiliation{Physics and Applied Mathematics Unit, Indian Statistical Institute, 203 Barrackpore Trunk Road, Kolkata-700 108, India}

\author{Sudin Ganguly}
\email{sudinganguly@gmail.com}
\affiliation{Department of Physics, Adamas University, Adamas Knowledge City, Barasat-Barrackpore Road, 24 Parganas North, Kolkata 700 126, India}

\author{Santanu K. Maiti}
\email{santanu.maiti@isical.ac.in}
\affiliation{Physics and Applied Mathematics Unit, Indian Statistical Institute, 203 Barrackpore Trunk Road, Kolkata-700 108, India}


\begin{abstract}
Using Hartree-Fock mean-field approach, we study the role of on-site ($U$) and extended ($V$) Hubbard interactions on the existence and evolution of edge modes in a half-filled Su-Schrieffer-Heeger (SSH) chain. We analyze the energy spectrum, local probability amplitudes, and site-resolved charge and spin density profiles across topological, critical, and trivial hopping regimes. We find that the localization of bound states is controlled by the ratio $2V/U$, with edge spin-density-wave modes for $U>2V$ and mid-chain charge-density-wave domain walls for $U<2V$, independent of band topology. These results establish the correlation-driven origin of localized states in finite one-dimensional chains.
\end{abstract}

\maketitle
{\it Introduction}. In systems with disorder, the typical outcome is localization. In contrast, topological systems support robust boundary-localized states protected by global properties of the bulk band structure. A paradigmatic example is the Su-Schrieffer-Heeger (SSH) model~\cite{ssh1,ssh2}, a one-dimensional dimerized lattice that hosts zero-energy edge modes in its topologically nontrivial phase, as dictated by the bulk-boundary correspondence~\cite{hasan,jkasb}. Owing to its simplicity, the SSH model serves as a minimal model for studying topological phenomena in low-dimensional systems. 

Most studies of the SSH model have focused on the non-interacting limit, where its topological properties are well understood in terms of band theory and symmetry considerations~\cite{ssh1,ssh2,jkasb,ryu}. The model has been extensively investigated across a wide range of platforms, including electronic systems~\cite{ssh1,ssh2,heeger}, photonic lattices~\cite{lu,ozawa}, cold atoms~\cite{atala,meier}, and mechanical metamaterials~\cite{huber,gma}. While the non-interacting framework has yielded significant insight, comparatively fewer studies have addressed electron-electron interactions, which can give rise to novel phenomena in topological systems. 

Interacting extensions of the SSH model have been studied within Hubbard-type and related interaction frameworks. The SSH model with on-site Hubbard interaction has been analyzed using the entanglement spectrum~\cite{bt} and correlation functions~\cite{barbiero} to characterize interaction effects on topology, while SSH chains with on-site interaction also exhibit interaction-driven topological phases~\cite{nhle}. Finite-filling regimes have also been explored, including studies incorporating both on-site and nearest-neighbor interactions in the presence of a magnetic field~\cite{mikhail}, as well as exact diagonalization analyses of interacting SSH chains~\cite{yxwang}. The evolution of edge states has further been examined in the presence of on-site Hubbard and Kondo interactions, demonstrating interaction-induced modifications of boundary modes~\cite{kumar}. In addition, bosonization studies of the interacting SSH model with nearest-neighbor Hubbard-type interactions provide a low-energy field-theoretic description of interaction effects and their impact on the topological phase~\cite{tjin}. However, a systematic understanding of the combined role of on-site and nearest-neighbor interactions remains relatively unexplored, particularly regarding their impact on localization and real-space properties.

In this work, we address this gap by studying an interacting SSH model with both on-site and nearest-neighbor interactions within a mean-field framework, focusing on the resulting spectral properties and real-space distribution of eigenstates.     We focus on the half-filled case, where each edge state is occupied by a single electron and therefore susceptible to spin polarization by $U$, and where the competition between SDW and CDW orderings is most sharply defined. Our analysis reveals a rich set of phenomena beyond the non-interacting picture, including unconventional boundary localization and interaction-induced mid-gap domain-wall states, highlighting the nontrivial role of interactions in driving localization beyond the non-interacting topological picture.

\begin{figure}[h]
    \centering
    \includegraphics[width=0.49\textwidth]{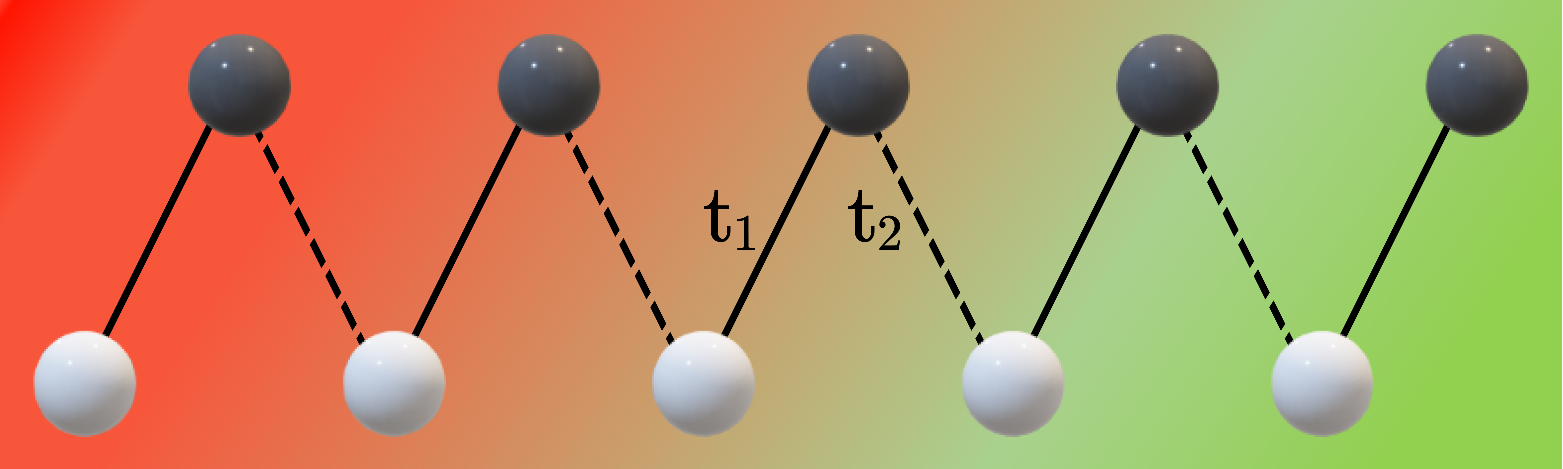}
    \caption{(Color online). Schematic diagram of the one-dimensional dimerized chain, where $t_1$ (solid line) and $t_2$ (dotted line) denote the intra- and inter-unit-cell hopping amplitudes, respectively. The white and black balls represent the two sublattices.}
    \label{setup}
\end{figure}

{\it Model}. The extended Hubbard Hamiltonian for the SSH model has the form~\cite{hubbard, hirsch, pgj}
\begin{eqnarray}
H &=&   \sum_{i,\sigma} \left(t+(-1)^i\delta t\right) \left(c_{i,\sigma}^\dagger c_{i+1,\sigma}+\text{h.c.}\right)\nonumber\\
&+&\sum_{i} U n_{i,\uparrow} n_{i,\downarrow} + \sum_{i} V n_{i} n_{i+1}.
\label{eqn1}
\end{eqnarray} 
Here, $c_{i,\sigma}^\dagger$ ($c_{i,\sigma}$) creates (annihilates) an electron with spin $\sigma$ at site $i$, and $n_i = n_{i,\uparrow} + n_{i,\downarrow}$ is the total number operator at site $i$.  The first term represents the nearest-neighbor hopping of the SSH model, where $t \pm \delta t$ are alternating hopping amplitudes (see Fig.~\ref{setup}). Denoting the intracell and intercell hopping amplitudes as $t_1$ and $t_2$, respectively, we have $t=\left(t_2 + t_1\right)/2$ and $\delta t=\left(t_2 - t_1\right)/2$. 
The system is in a topologically nontrivial phase for $t_1 < t_2$ and in a trivial phase for $t_1 > t_2$. The second term corresponds to the on-site Hubbard interaction between electrons of opposite spin at the same site, with $U$ denoting the interaction strength. The third term describes the nearest-neighbor density-density (extended Hubbard) interaction, where $V$ is the corresponding interaction strength.

To determine the energy eigenvalues of the interacting system described by Eq.~(\ref{eqn1}), we employ the generalized Hartree-Fock (mean-field) approximation, under which the interaction terms are decoupled as
\begin{equation}
n_{i,\uparrow} n_{i,\downarrow} \approx 
\langle n_{i,\uparrow} \rangle n_{i,\downarrow} +
\langle n_{i,\downarrow} \rangle n_{i,\uparrow}
- \langle n_{i,\uparrow} \rangle \langle n_{i,\downarrow} \rangle,
\end{equation}
and
\begin{equation}
n_i n_{i+1} \approx 
\langle n_i \rangle n_{i+1} + \langle n_{i+1} \rangle n_i 
- \langle n_i \rangle \langle n_{i+1} \rangle.
\end{equation}
Within this approximation, the Hamiltonian decouples into two independent parts corresponding to up- and down-spin electrons with renormalized site energies. 
The full Hamiltonian in Eq.~(\ref{eqn1}) can thus be written in the decoupled form as
\begin{eqnarray}
H =  H_\uparrow + H_\downarrow - U\sum_{i} \langle n_{i,\uparrow}\rangle \langle n_{i,\downarrow}\rangle
+ V \sum_{i} \langle n_{i}\rangle \langle n_{i+1}\rangle,
\label{mft-ham}
\end{eqnarray}
where the last two terms represent a constant energy shift to the total energy. The decoupled Hamiltonians $H_\uparrow$ and $H_\downarrow$ are,
\begin{eqnarray}
H_{\uparrow} 
&=& \sum_{i=1} \epsilon_{i,\uparrow}^{\text{eff}} 
   c_{i,\uparrow}^\dagger c_{i,\uparrow} \nonumber\\
&+& \sum_{i} \left(t+(-1)^i\delta t\right) 
   \left( c_{i,\uparrow}^\dagger 
   c_{i+1,\uparrow} + \text{h.c.} \right),\\
\label{eqn4a}
\text{and,}\nonumber\\
H_{\downarrow} 
&=& \sum_{i=1} \epsilon_{i,\downarrow}^{\text{eff}} 
   c_{i,\downarrow}^\dagger c_{i,\downarrow} \nonumber\\
&+& \sum_{i=1} \left(t+(-1)^i\delta t\right) 
   \left( c_{i,\downarrow}^\dagger 
   c_{i+1,\downarrow} + \text{h.c.} \right),
\label{eqn4b}
\end{eqnarray} 
with the renormalized site energies,
\begin{eqnarray}
\epsilon_{i,\uparrow}^{\text{eff}} &=&  U \langle n_{i,\downarrow} \rangle + V  \left(\langle n_{i+1} \rangle + \langle n_{i-1} \rangle\right),\\
\epsilon_{i,\downarrow}^{\text{eff}} &=&  U \langle n_{i,\uparrow} \rangle + V  \left(\langle n_{i+1} \rangle + \langle n_{i-1} \rangle\right).
\label{eqn3}
\end{eqnarray}

The energy eigenvalues are obtained using a self-consistent mean-field procedure. Starting with initial guess values of $\langle n_{i,\uparrow} \rangle$ and $\langle n_{i,\downarrow} \rangle$, the decoupled Hamiltonians $H_{\uparrow}$ and $H_{\downarrow}$ are constructed and diagonalized to obtain the eigenvalues and eigenvectors. For the half-filled case, the local spin densities are updated by summing over the lowest occupied states in each spin channel. This procedure is repeated iteratively until convergence is achieved, yielding the final self-consistent energy spectrum and eigenstates.

{\it Discussion}. The numerical results are obtained for a half-filled chain consisting of 100 sites, and we present the results only for the spin-resolved mean-field Hamiltonian $H_{\uparrow}$. Since both spin sectors are identical, the down-spin Hamiltonian does not require separate discussion. We set intercell hopping $t_2=1\,$eV as the reference energy scale. For the topological phase, we take $t_1 = 0.5\,$eV while for the trivial phase, we take $t_1 = 1.5\,$eV.

In a one-dimensional half-filled chain, electron-electron interactions can induce spin-density-wave (SDW) and charge-density-wave (CDW) phases. The on-site Hubbard interaction $U$ suppresses double occupancy and favors antiferromagnetic spin ordering, leading to an SDW or Mott-insulating phase~\cite{hubbard}. In contrast, the nearest-neighbor interaction $V$ promotes alternating charge occupation, resulting in a CDW phase~\cite{hirsch}. When both $U$ and $V$ are present, they compete with each other. The SDW phase is favored for $U > 2V$, while the CDW phase becomes stable for $U < 2V$~\cite{hirsch}. This interplay becomes particularly important in the SSH chain, where electron correlation and topology coexist. We therefore examine how $U$ and $V$ affect edge-state localization and stability in both topological, critical, and trivial phases, along with the emergence of CDW and SDW orderings.

{\it Topological phase}. We first examine the energy spectrum of the SSH chain in the topological phase with $t_1 = 0.5$ 
\begin{figure}[ht]
    \includegraphics[width=0.48\textwidth]{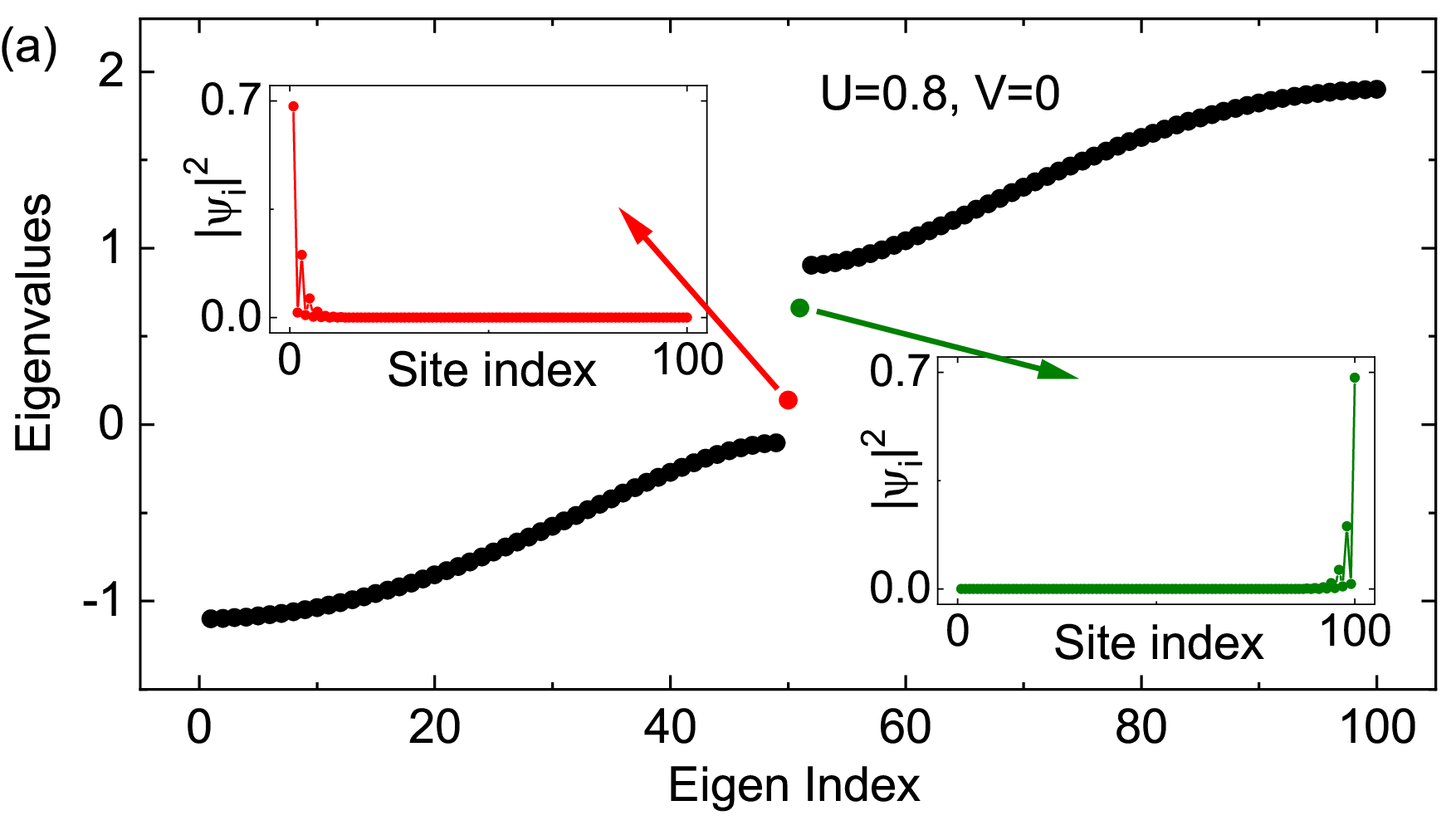}\hfill\vskip 0.1 in
    \includegraphics[width=0.24\textwidth]{fig2b}\hfill\includegraphics[width=0.24\textwidth]{fig2c}
    \caption{(Color online). (a) Energy eigenvalue spectrum as a function of eigenstate index for a 100-site half-filled SSH chain with $t_1=0.5$ and $t_2=1$, in the presence of on-site Hubbard interaction $U=0.8$. The two in-gap states are marked by red and green dots. The insets show the corresponding local probability amplitudes $\left|\psi_i\right|^2$ for these two states, represented by the red and green curves, respectively, confirming their edge-localized nature. Site-resolved (b) charge density $n_i = n_{i,\uparrow}+n_{i,\downarrow}$ and (c) spin density $m_i = n_{i,\uparrow}-n_{i,\downarrow}$ plotted as a function of site index. The alternating red and blue dots in (b) and (c) distinguish adjacent site indices.}
    \label{topo1}  
\end{figure}    
in the presence of only on-site Hubbard interaction, as shown in Fig.~\ref{topo1}(a), where the energy eigenvalues are plotted as a function of eigenvalue index. Here the strength of on-site Hubbard interaction is fixed at $U=0.8$. The spectrum clearly exhibits a finite energy gap with only two states lying inside the gap. These two in-gap states are marked by red and green dots for better visibility. The insets show the corresponding local probability amplitudes, $\left\lvert\psi_i\right\rvert^2$, of these two states. The red curve in the left inset corresponds to the red dot in the spectrum, while the green curve in the right inset corresponds to the green dot. Both states are strongly localized at the two ends of the chain, confirming their edge-state nature. This demonstrates that the topological boundary modes of the SSH chain persist under finite on-site interaction.

To further understand the role of electronic correlation, we analyze the CDW and SDW orderings, characterized by $n_i \left(=n_{i,\uparrow}+n_{i,\downarrow}\right)$ and $m_i \left(= n_{i,\uparrow}-n_{i,\downarrow}\right)$, respectively. The site-resolved charge density and spin density are plotted as a function of site index in Figs.~\ref{topo1}(b) and (c), respectively. The CDW profile is completely flat, with $n_i = 1$ across all sites, confirming the absence of any charge order. In contrast, the spin density shows strong alternating values localized near the edges ($\lvert m_i\rvert \approx 0.85$), which decay exponentially into a nearly uniform bulk. This SDW behavior arises from the combined effect of on-site Hubbard interaction and SSH dimerization. As confirmed by the in-gap states in Fig.~\ref{topo1}(a), the dimerized hopping produces boundary modes localized at the edges of the chain. Under finite $U$, these edge modes develop a spin imbalance to minimize the on-site Coulomb repulsion, resulting in enhanced edge magnetization. In a uniform hopping chain, the on-site Hubbard interaction generally induces staggered spin ordering across the entire chain~\cite{hubbard}. Here, however, the spin density accumulates only near the boundaries because the underlying electronic states are themselves edge-localized. Therefore, the observed edge-localized SDW order is simply the spin-polarized manifestation of the topological boundary modes under finite Hubbard interaction.

Further increasing $U$, the edge states eventually disappear. Although not shown here for brevity, increasing $U$ gradually enhances the SDW ordering, leading to a more uniform spin-density modulation across the chain. Concurrently, the bulk energy gap increases with $U$. The corresponding in-gap states remain non-degenerate but progressively shift towards the bulk continuum. As a result, their edge localization becomes significantly weaker, and around $U \approx 1.7$, they merge into the bulk spectrum, thereby losing their distinct edge-state character. This particular feature is discussed later in this work.

\begin{figure}[ht]
    \includegraphics[width=0.48\textwidth]{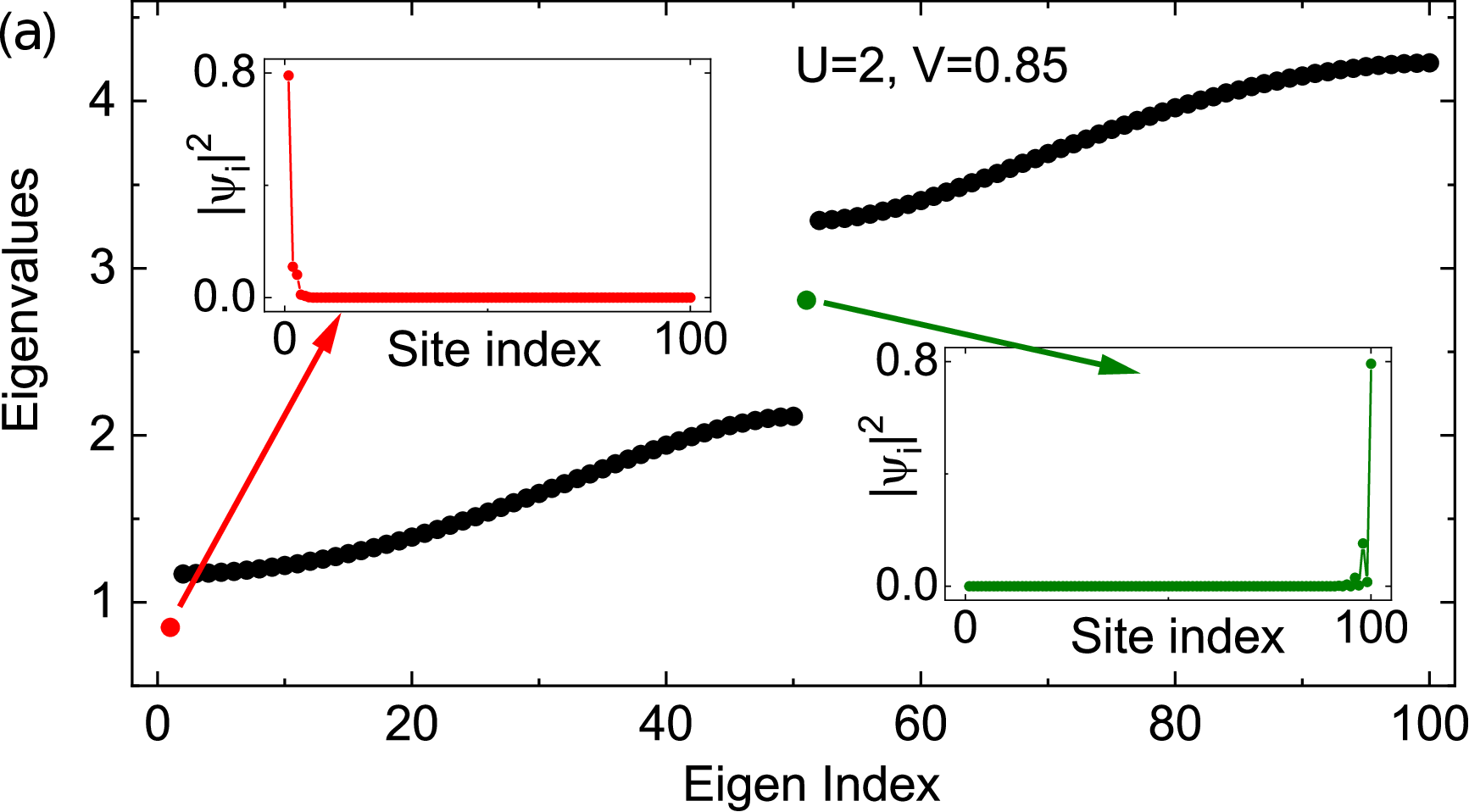}\hfill\vskip 0.1 in
    \includegraphics[width=0.24\textwidth]{fig3b}\hfill\includegraphics[width=0.24\textwidth]{fig3c}
    \caption{(Color online). (a) Energy eigenvalue spectrum as a function of eigenstate index for a 100-site half-filled SSH chain with $t_1=0.5$ and $t_2=1$, in the presence of both on-site and nearest-neighbor Hubbard interactions $U=2$ and $V=0.85$. The isolated state below the lower band is marked by a red dot, and the in-gap state is marked by a green dot. The insets show the corresponding local probability amplitudes $\left|\psi_i\right|^2$ for these two states, represented by the red and green curves, respectively. Site-resolved (b) charge density $n_i$ and (c) spin density $m_i$ plotted as a function of site index. The alternating red and blue dots in (b) and (c) distinguish adjacent site indices.}
    \label{topo2}  
\end{figure}

We now examine the combined effect of on-site and nearest-neighbor interactions. Figure~\ref{topo2}(a) shows the energy spectrum for $U=2$ and $V=0.85$. The spectrum exhibits one isolated state below the lower band denoted by red dot and one state inside the bulk gap denoted by green dot. The insets display the local probability density $\left\lvert\psi_i\right\rvert^2$ of these two states, both of which are strongly localized at the edges of the SSH chain, confirming their edge-state character. This demonstrates that the nearest-neighbor interaction $V$ restores the topological boundary modes at $U=2$, where pure on-site repulsion alone would have driven them into the bulk continuum (as discussed earlier, the in-gap states disappear around $U\approx 1.7$ for $V=0$).

The site-resolved charge density $n_i$ and spin density $m_i$ are plotted in Figs.~\ref{topo2}(b) and (c), respectively. The CDW profile shows weak Friedel oscillations near the edges~\cite{brendel}, deviating from $n_i=1$ by about $\pm 10\%$ before decaying into a uniform bulk over roughly 10-15 sites. No long-range charge order develops. This is consistent with the condition $U>2V$ ($U=2$, $V=0.85$), which favors SDW over CDW in the extended Hubbard model. The SDW profile reveals a staggered magnetization that extends across the entire chain, with a bulk value $\lvert m_i\rvert\approx 0.3$ and markedly enhanced edge moments reaching $\lvert m_i\rvert\approx 0.85$. This behavior originates from the restored topological boundary modes. At the edges, where the boundary modes have large probability amplitude, the on-site Hubbard interaction $U$ induces strong spin polarization. The nearest-neighbor interaction $V$ further stabilizes the bulk SDW order, consistent with $U>2V$, while the boundary modes, having large probability amplitude at the edges, acquire enhanced spin polarization on top of the bulk antiferromagnetic background. In the charge sector, the boundary modes produce the observed Friedel oscillations in $n_i$ near the edges, while the bulk remains uniform. Thus, the CDW and SDW profiles are direct consequences of the edge states restored by $V$, where the edge modes give rise to both the charge modulations and the spin order, with $V$ extending the latter across the entire chain. This is in contrast to the weak-$U$ case where edge magnetism exists without bulk order, highlighting the role of $V$ in establishing a coexisting bulk and edge antiferromagnetic phase.

Next we examine the case $U=0.8$ and $V=0.6$. Here $U<2V$, a regime where the extended 
\begin{figure}[ht]
    \includegraphics[width=0.48\textwidth]{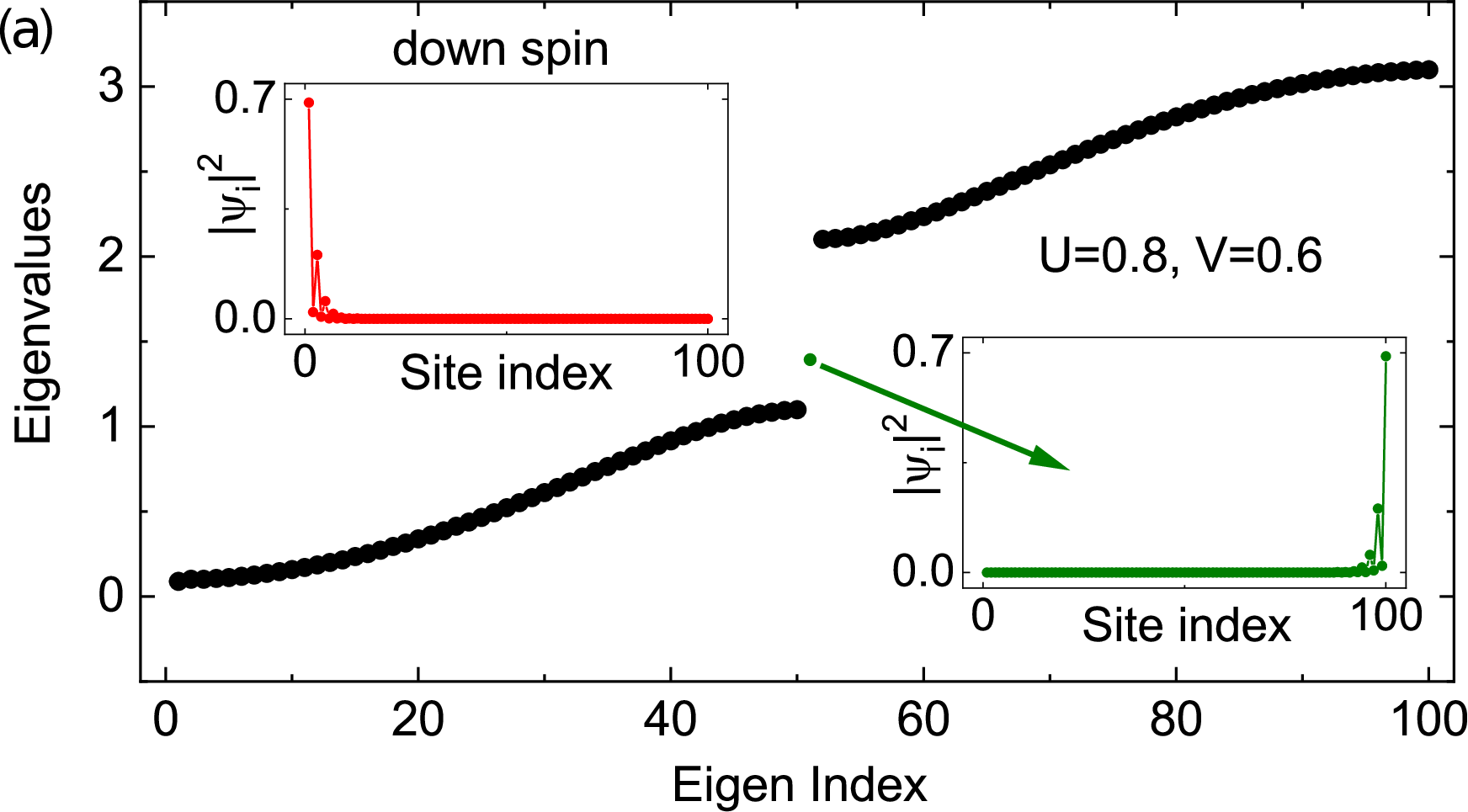}\hfill\vskip 0.1 in
    \includegraphics[width=0.24\textwidth]{fig4b}\hfill\includegraphics[width=0.24\textwidth]{fig4c}
    \caption{(Color online). (a) Energy eigenvalue spectrum as a function of eigenstate index for a 100-site half-filled SSH chain with $t_1=0.5$ and $t_2=1$, in the presence of both on-site and nearest-neighbor Hubbard interactions $U=0.8$ and $V=0.6$. The single in-gap state is marked by a green dot. The right inset shows the local probability amplitude $\left|\psi_i\right|^2$ for the spin-up case (green curve), while the right inset shows the same for the spin-down case (red curve). Site-resolved (b) charge density $n_i$ and (c) spin density $m_i$ plotted as a function of site index. The alternating red and blue dots in (b) and (c) distinguish adjacent site indices.}
    \label{topo3}  
\end{figure} 
Hubbard model favors CDW over SDW order in the bulk. Figure~\ref{topo3}(a) shows the energy spectrum, which exhibits only a single in-gap state denoted by a green dot. The insets reveal a striking asymmetry. For the spin-up sector, the local probability density is localized exclusively at the left edge (right inset with green curve), while for the spin-down sector, it is localized at the right edge (left inset with red curve). The two spin sectors thus host boundary modes at opposite ends of the chain.

The site-resolved charge density and spin density are plotted in Figs.~\ref{topo3}(b) and (c), respectively. The CDW profile shows weak Friedel oscillations near the edges~\cite{brendel}, deviating from $n_i=1$ by about $\pm10\%$ before decaying into a uniform bulk. The SDW profile reveals a large positive spin density at one edge ($m_i\approx 0.7$) and a large negative spin density at the other ($m_i\approx -0.7$), with strong alternating values that decay exponentially into a nearly non-magnetic bulk. This behavior directly reflects the spin-resolved edge localization. The spin-up state localized at the right edge produces a net positive spin density there, while the spin-down state localized at the left edge produces a net negative spin density. The bulk remains unpolarized, consistent with $U<2V$ case, where CDW correlations are expected to dominate and SDW order is suppressed. This spin-dependent spatial separation of boundary modes is a distinct consequence of the nearest-neighbor interaction $V$, with no analog in the pure-$U$ or pure SSH limits.

\begin{figure}[ht]
    \includegraphics[width=0.48\textwidth]{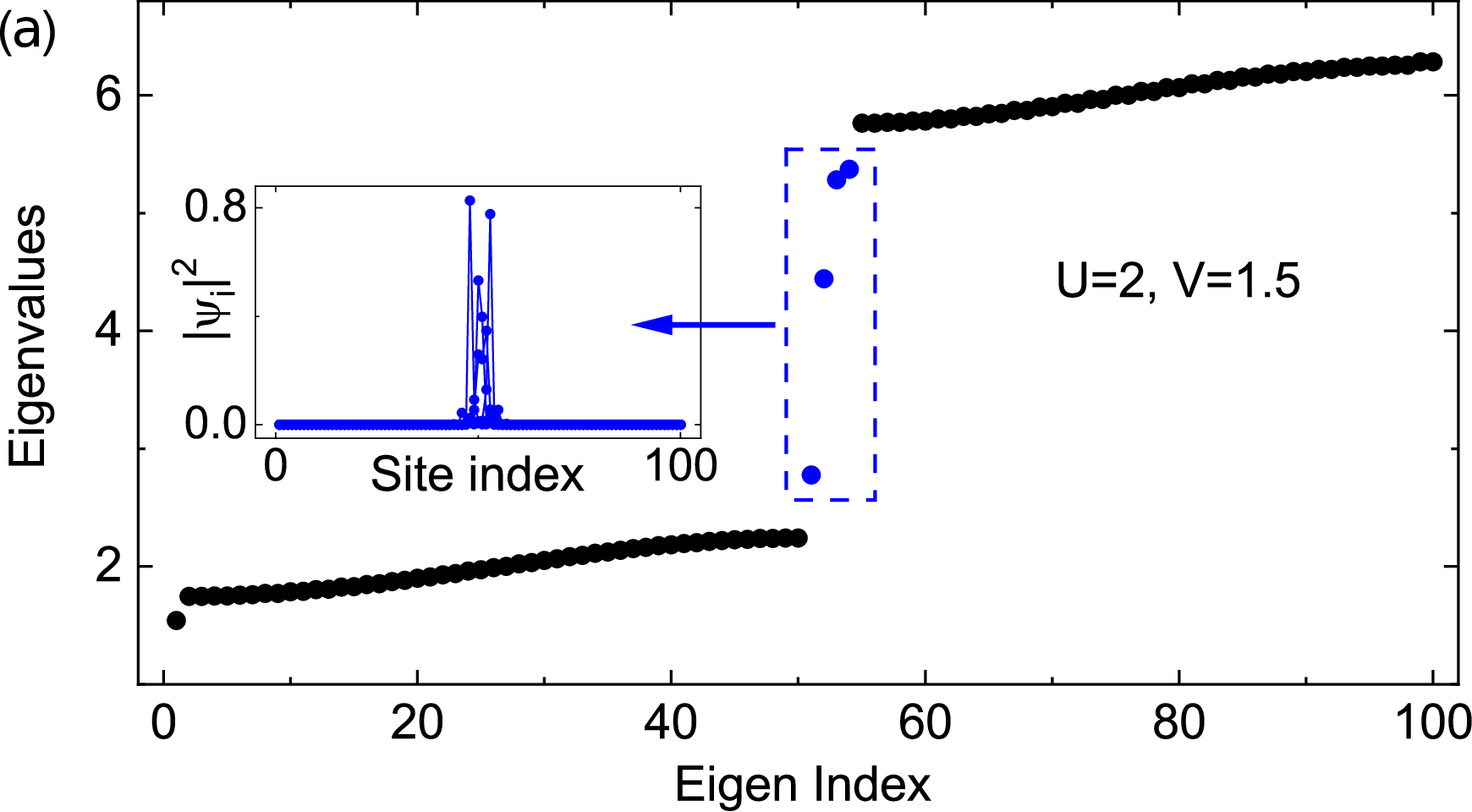}\hfill\vskip 0.1 in
    \includegraphics[width=0.24\textwidth]{fig5b}\hfill\includegraphics[width=0.24\textwidth]{fig5c}
    \caption{(Color online). (a) Energy eigenvalue spectrum as a function of eigenstate index for a 100-site half-filled SSH chain with $t_1=0.5$ and $t_2=1$, in the presence of both on-site and nearest-neighbor Hubbard interactions $U=2$ and $V=1.5$. The in-gap states are marked by blue dots. The inset shows the corresponding local probability amplitudes $\left\lvert\psi_i\right\rvert^2$. Site-resolved (b) charge density $n_i$ and (c) spin density $m_i$ plotted as a function of site index. The alternating red and blue dots in (b) and (c) distinguish adjacent site indices.}
    \label{topo4}  
\end{figure}

We now consider $U=2$ and $V=1.5$, where $U<2V$ strongly favors CDW order. The energy spectrum (Fig.~\ref{topo4}(a)) shows multiple in-gap states marked by blue dots. Unlike the edge-dominated cases, the local probability density reveals these states are localized at the center of the chain. The site-resolved charge density (Fig.~\ref{topo4}(b)) 
explains this shift. A robust CDW pattern extends across the chain, but with opposite phasing in the two halves. The first half follows a high-low-high-low pattern on odd-even sites, while the second half reverses to low-high-low-high. This creates a domain wall precisely at the chain center, where the in-gap states reside. The SDW profile is vanishingly small everywhere as observed in Fig.~\ref{topo4}(c), consistent with the condition $U<2V$. These mid-chain states are domain-wall bound modes, well-known in the 1D extended Hubbard model. For instance, Tomita and Fukutome~\cite{tomita} have shown that SDW-CDW domain walls host fractionalized in-gap states in the one-dimensional extended Hubbard model, while Matsuno et al.~\cite{matsuno} and Ohki et al.~\cite{ohki} demonstrated that domain walls between charge-ordered regions with opposite polarization host bound states in related extended Hubbard systems.  Thus, increasing $V$ shifts the in-gap states from the edges to the center of the chain, reflecting the transition from a boundary-localized regime to a charge-ordered phase dominated by $V$.

\begin{figure}[ht]
    \includegraphics[width=0.48\textwidth]{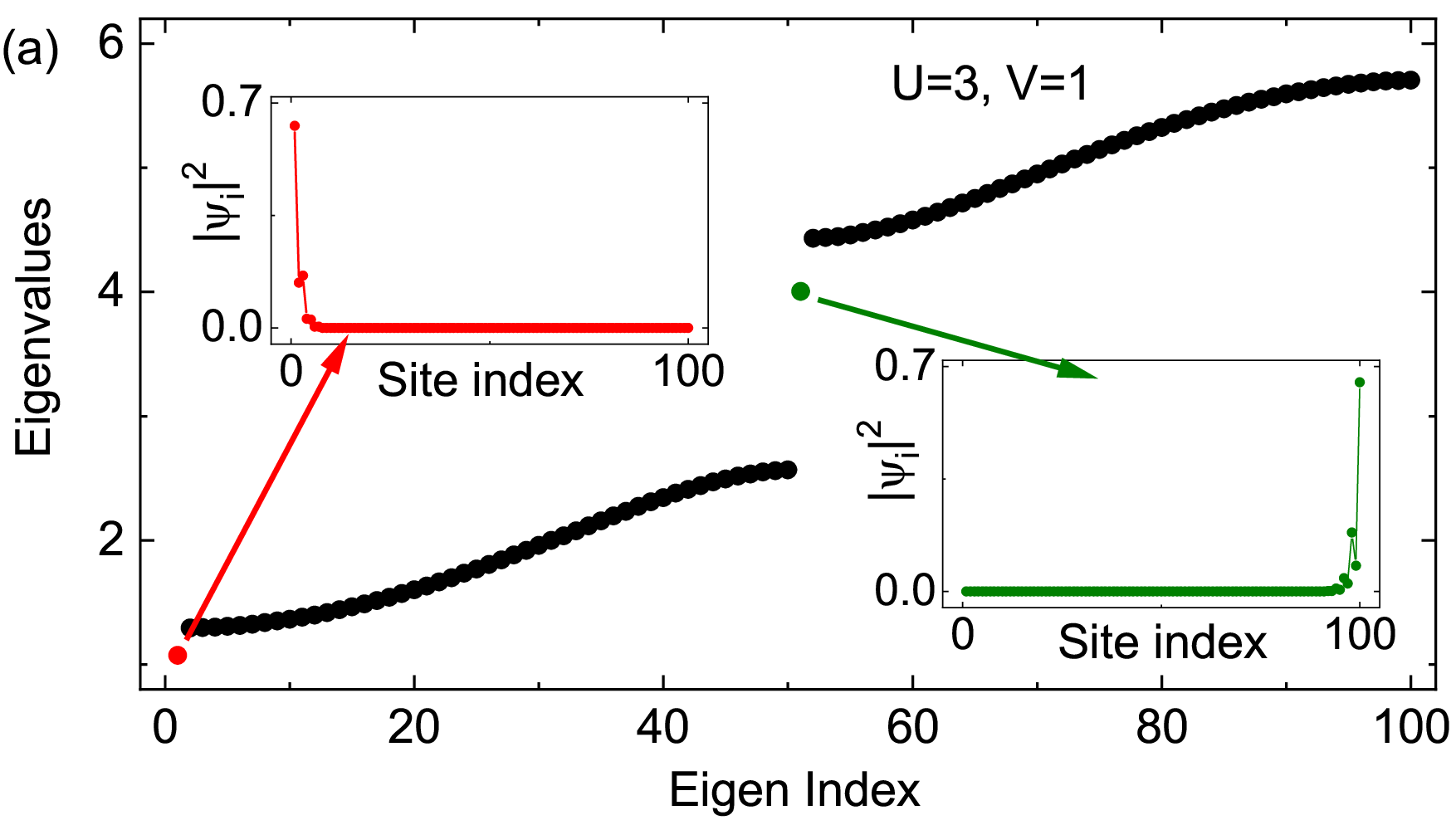}\hfill\vskip 0.1 in
    \includegraphics[width=0.24\textwidth]{fig6b}\hfill\includegraphics[width=0.24\textwidth]{fig6c}
    \caption{(Color online). (a) Energy eigenvalue spectrum as a function of eigenstate index for a 100-site half-filled SSH chain with $t_1=1$ and $t_2=1$, in the presence of both on-site and nearest-neighbor Hubbard interactions $U=3$ and $V=1$. The isolated state below the lower band is marked by a red dot, and the in-gap state is marked by a green dot. The insets show the corresponding local probability amplitudes $\left|\psi_i\right|^2$ for these two states, represented by the red and green curves, respectively. Site-resolved (b) charge density $n_i$ and (c) spin density $m_i$ plotted as a function of site index. The alternating red and blue dots in (b) and (c) distinguish adjacent site indices.}
    \label{critical1}  
\end{figure}
The results presented so far were obtained for the dimerized chain with $t_1=0.5$, i.e., in the topologically nontrivial phase of the SSH model. To determine whether these states are intrinsic to the SSH topology or arise more generally from interactions and open boundaries, we now examine the critical point $(t_1=1)$ and the trivial phase $(t_1=1.5)$. We note that in both regimes, pure on-site $U$ alone does not produce edge-localized in-gap states, in contrast to the topological phase at weak $U$.

{\it Critical point}. We first examine the critical point $t_1=1$. Figure~\ref{critical1}(a) shows the energy 
spectrum for $U=3$ and $V=1$. Since $U>2V$, SDW correlations dominate. The spectrum exhibits one isolated state below the lower band denoted by a red dot and one state inside the bulk gap denoted by a green dot. The local probability density, as observed from the insets, confirms they are localized at the edges. The CDW profile (Fig.~\ref{critical1}(b)) shows weak Friedel oscillations, while the SDW profile (Fig.~\ref{critical1}(c)) displays enhanced edge magnetization, consistent with the behavior observed in the topological phase (Fig.~\ref{topo2}). We note that, unlike the topological phase where edge modes exist even in the non-interacting limit, the critical point hosts no such boundary modes at $U=V=0$. The edge-localized states here are therefore purely interaction-driven, arising from the combined effect of $U$ and $V$ under open boundary conditions.

We now consider $U=3$ and $V=2$ at the critical point $t_1=1$. Here $U<2V$ favors CDW order. 
\begin{figure}[ht]
    \includegraphics[width=0.48\textwidth]{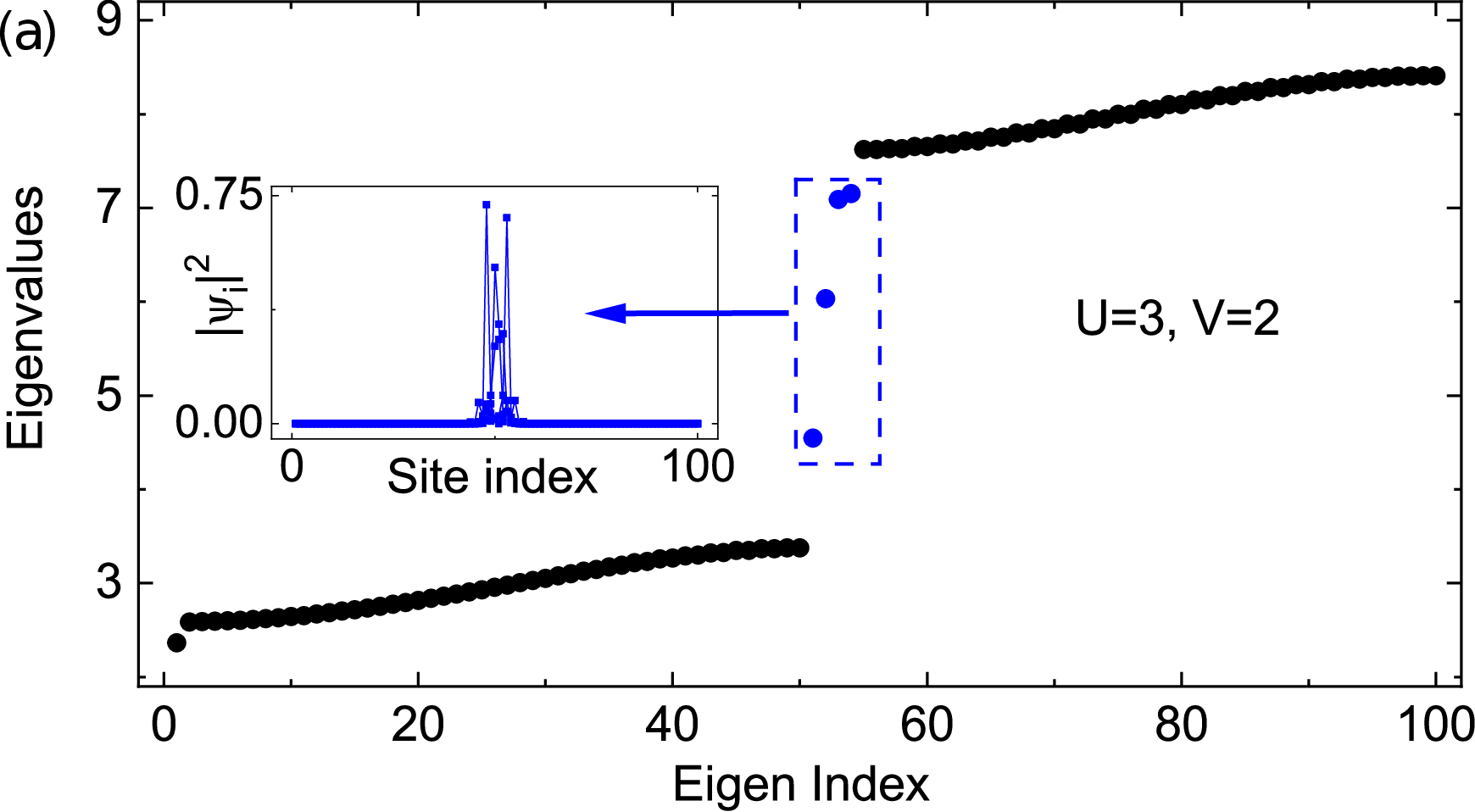}\hfill\vskip 0.1 in
    \includegraphics[width=0.24\textwidth]{fig7b}\hfill\includegraphics[width=0.24\textwidth]{fig7c}
    \caption{(Color online). (a) Energy eigenvalue spectrum as a function of eigenstate index for a 100-site half-filled SSH chain with $t_1=1$ and $t_2=1$, in the presence of both on-site and nearest-neighbor Hubbard interactions $U=3$ and $V=2$. The in-gap states are marked by blue dots. The inset shows the corresponding local probability amplitudes $\left\lvert\psi_i\right\rvert^2$. Site-resolved (b) charge density $n_i$ and (c) spin density $m_i$ plotted as a function of site index. The alternating red and blue dots in (b) and (c) distinguish adjacent site indices.}
    \label{critical2}  
\end{figure} 
The energy spectrum (Fig.~\ref{critical2}(a)) shows multiple in-gap states localized at the center of the chain. The charge density (Fig.~\ref{critical2}(b)) reveals a CDW domain wall with opposite phasing in the two halves, identical to the pattern observed in the topological phase. The SDW profile (Fig.~\ref{critical2}(c)) is vanishingly small throughout the chain, except near the domain wall where noticeable fluctuations appear, consistent with the suppression of SDW order in the $U<2V$ regime. Thus, as in the topological phase, increasing $V$ shifts the in-gap states from the edges to the center, reflecting the dominance of CDW order.

\begin{figure}[ht]
    \includegraphics[width=0.48\textwidth]{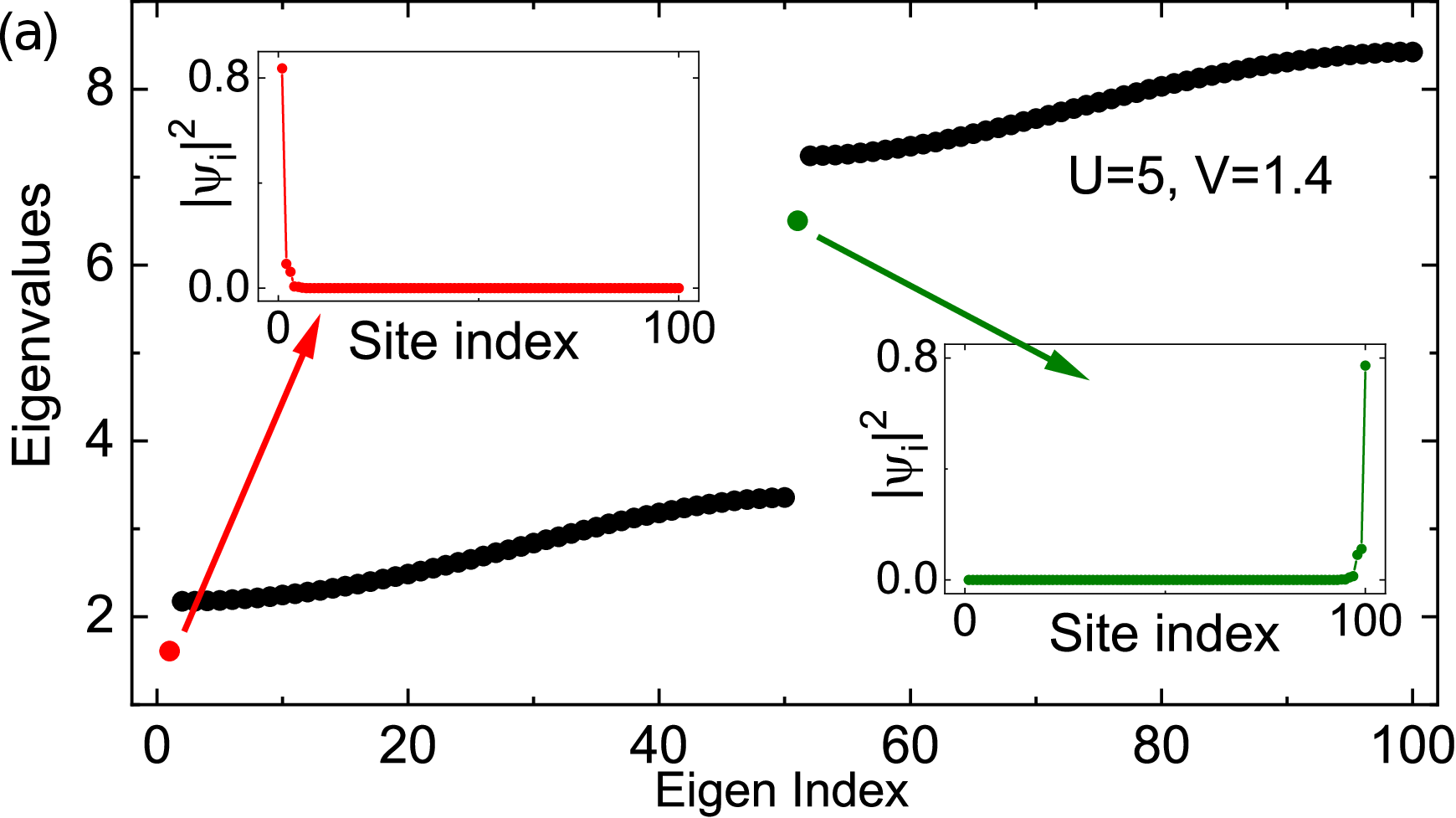}\hfill\vskip 0.1 in
    \includegraphics[width=0.24\textwidth]{fig8b}\hfill\includegraphics[width=0.24\textwidth]{fig8c}
    \caption{(Color online). (a) Energy eigenvalue spectrum as a function of eigenstate index for a 100-site half-filled SSH chain with $t_1=1.5$ and $t_2=1$, in the presence of both on-site and nearest-neighbor Hubbard interactions $U=5$ and $V=1.4$. The isolated state below the lower band is marked by a red dot, and the in-gap state is marked by a green dot. The insets show the corresponding local probability amplitudes $\left\vert\psi_i\right\vert^2$ for these two states, represented by the red and green curves, respectively. Site-resolved (b) charge density $n_i$ and (c) spin density $m_i$ plotted as a function of site index. The alternating red and blue dots in (b) and (c) distinguish adjacent site indices.}
    \label{trivial1}  
\end{figure}

{\it Trivial phase}. Finally, we examine the trivial phase by considering $t_1=1.5$. Figure~\ref{trivial1}(a) shows the energy spectrum for $U=5$ and $V=1.4$. Since $U>2V$, SDW correlations dominate. The spectrum exhibits in-gap states, and the local probability density confirms they are localized at the edges. As mentioned earlier, these edge states are also purely interaction-driven, arising from the combined effect of $U$ and $V$ in the presence of open boundaries. The CDW profile (Fig.~\ref{trivial1}(b)) shows weak Friedel oscillations near the edges, while the SDW profile (Fig.~\ref{trivial1}(c)) reveals staggered magnetization. This behavior is consistent with the topological phase and the critical point at similar $U>2V$ (e.g., Figs.~\ref{topo2} and ~\ref{critical1}), confirming that the edge-localized SDW physics is controlled by the interaction ratio rather than the underlying band topology.

\begin{figure}[ht]
    \includegraphics[width=0.48\textwidth]{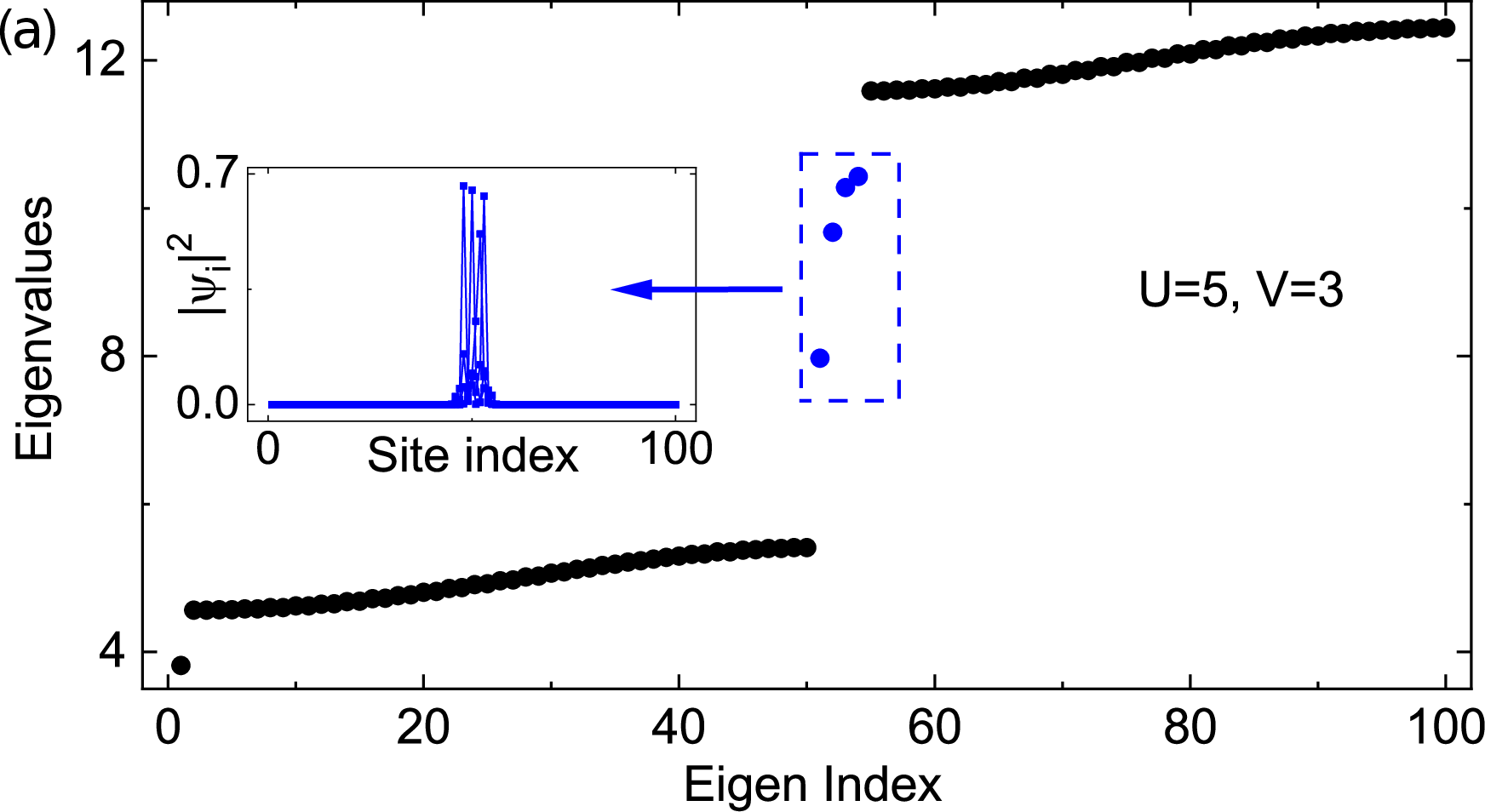}\hfill\vskip 0.1 in
    \includegraphics[width=0.24\textwidth]{fig9b}\hfill\includegraphics[width=0.24\textwidth]{fig9c}
    \caption{(Color online). (a) Energy eigenvalue spectrum as a function of eigenstate index for a 100-site half-filled SSH chain with $t_1=1.5$ and $t_2=1$, in the presence of both on-site and nearest-neighbor Hubbard interactions $U=5$ and $V=3$. The in-gap states are marked by blue dots. The inset shows the corresponding local probability amplitudes $\left\lvert\psi_i\right\rvert^2$. Site-resolved (b) charge density $n_i$ and (c) spin density $m_i$ plotted as a function of site index. The alternating red and blue dots in (b) and (c) distinguish adjacent site indices.}
    \label{trivial2}  
\end{figure}

We now consider $U=5$ and $V=3$ in the trivial phase with $t_1=1.5$. Here $U<2V$ favors CDW order. The energy spectrum (Fig.~\ref{trivial2}(a)) shows multiple in-gap states localized at the center of the chain. The charge density (Fig.~\ref{trivial2}(b)) reveals a CDW domain wall with opposite phasing in the two halves, identical to the pattern observed in the topological phase (Fig.~\ref{topo4}). The SDW profile (Fig.~\ref{trivial2}(c)) is vanishingly small throughout the chain, with noticeable fluctuations around the domain wall, consistent with $U<2V$. The mid-chain domain-wall states here are purely interaction-driven. Thus, as in the topological phase, the condition $U<2V$ shifts the in-gap states to the center, reflecting the dominance of CDW order independent of the underlying band topology.

The cases summarized in Table~\ref{table1} establish a clear pattern. For $U>2V$, localized states appear at the edges with enhanced SDW order, while for $U<2V$, they shift to the mid-chain and are associated with a CDW domain wall. Depending on the parameter regime, these states may lie inside the bulk gap or below the lower band. This behavior is observed across the topological, critical, and trivial hopping regimes, suggesting that the interaction ratio $2V/U$, rather than the SSH band topology, governs the spatial character of these localized states. To examine how this edge localization evolves continuously across the interaction parameter space, we now compute the maximum edge probability as a function of $U$ and $V$ for different hopping regimes.
\begin{table*}[htb]
\centering
\caption{Summary of localized state character and order profiles across hopping regimes. 
The condition $U > 2V$ favors SDW order with edge localization, while $U < 2V$ favors CDW order with mid-chain domain-wall states.}
\label{table1}
\renewcommand{\arraystretch}{1.3}
\begin{tabular}{|c|c|c|c|c|c|}
\hline
\multirow{2}{*}{Regime} & \multirow{2}{*}{$(U,V)$} & \multirow{2}{*}{Condition} & \multirow{2}{*}{Localization} & \multicolumn{2}{c|}{Order Profile} \\
\cline{5-6}
& & & & CDW & SDW \\
\hline
\multirow{4}{*}{Topological ($t_1=0.5$ and $t_2=1$)}
& $(0.8, 0)$  & --- & Edge & Flat & Edge-enhanced \\
\cline{2-6}
& $(2, 0.85)$ & $U > 2V$ & Edge & Friedel oscillations & Bulk + edge \\
\cline{2-6}
& $(0.8, 0.6)$ & $U < 2V$ & Edge (spin-split) & Friedel oscillations & Opposite edge signs \\
\cline{2-6}
& $(2, 1.5)$ & $U < 2V$ & Mid-chain (domain wall) & CDW domain wall & Vanishing \\
\hline
\multirow{2}{*}{Critical ($t_1=1$ and $t_2=1$)}
& $(3, 1)$ & $U > 2V$ & Edge & Friedel oscillations & Edge-enhanced \\
\cline{2-6}
& $(3, 2)$ & $U < 2V$ & Mid-chain (domain wall) & CDW domain wall & Vanishing$^\dagger$ \\
\hline
\multirow{2}{*}{Trivial ($t_1=1.5$ and $t_2=1$)}
& $(5, 1.4)$ & $U > 2V$ & Edge & Friedel oscillations & Edge-enhanced \\
\cline{2-6}
& $(5, 3)$ & $U < 2V$ & Mid-chain (domain wall) & CDW domain wall & Vanishing$^\dagger$ \\
\hline
\end{tabular}

\vspace{0.1in}
$^\dagger$Vanishing except for noticeable fluctuations around the domain wall.

\end{table*}

\begin{figure}[ht]
\includegraphics[width=0.24\textwidth]{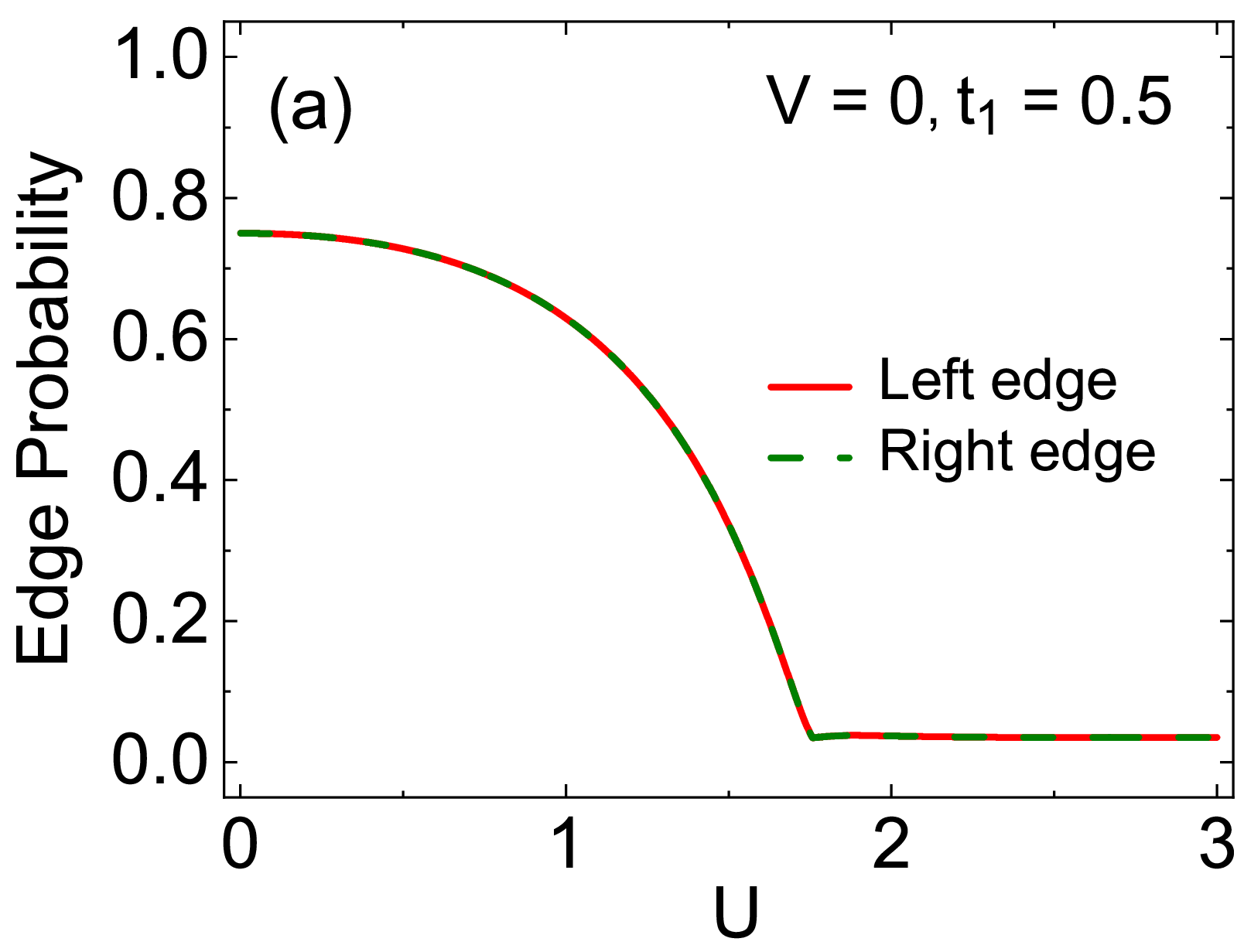}\hfill\includegraphics[width=0.24\textwidth]{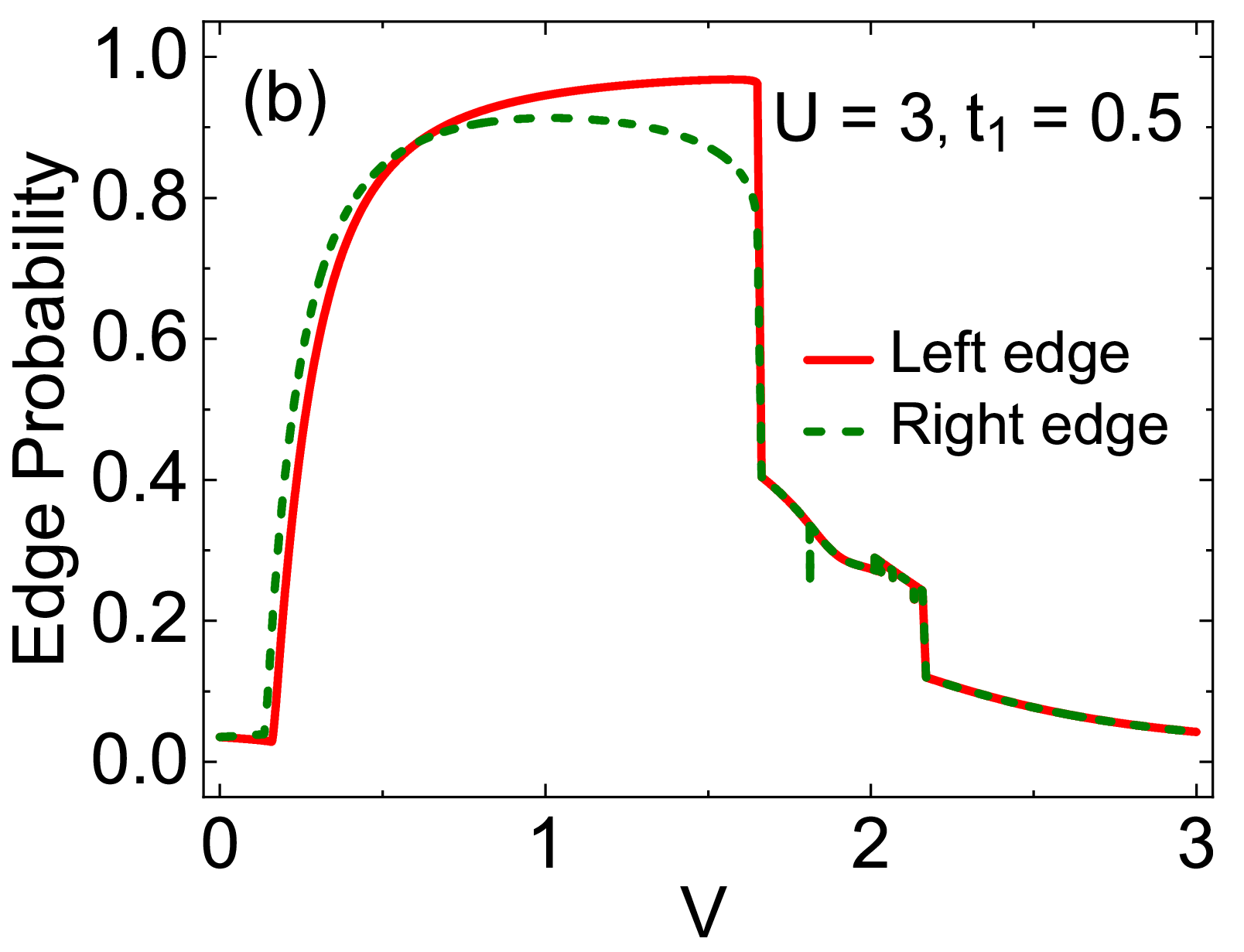}\vskip 0.1 in
\includegraphics[width=0.24\textwidth]{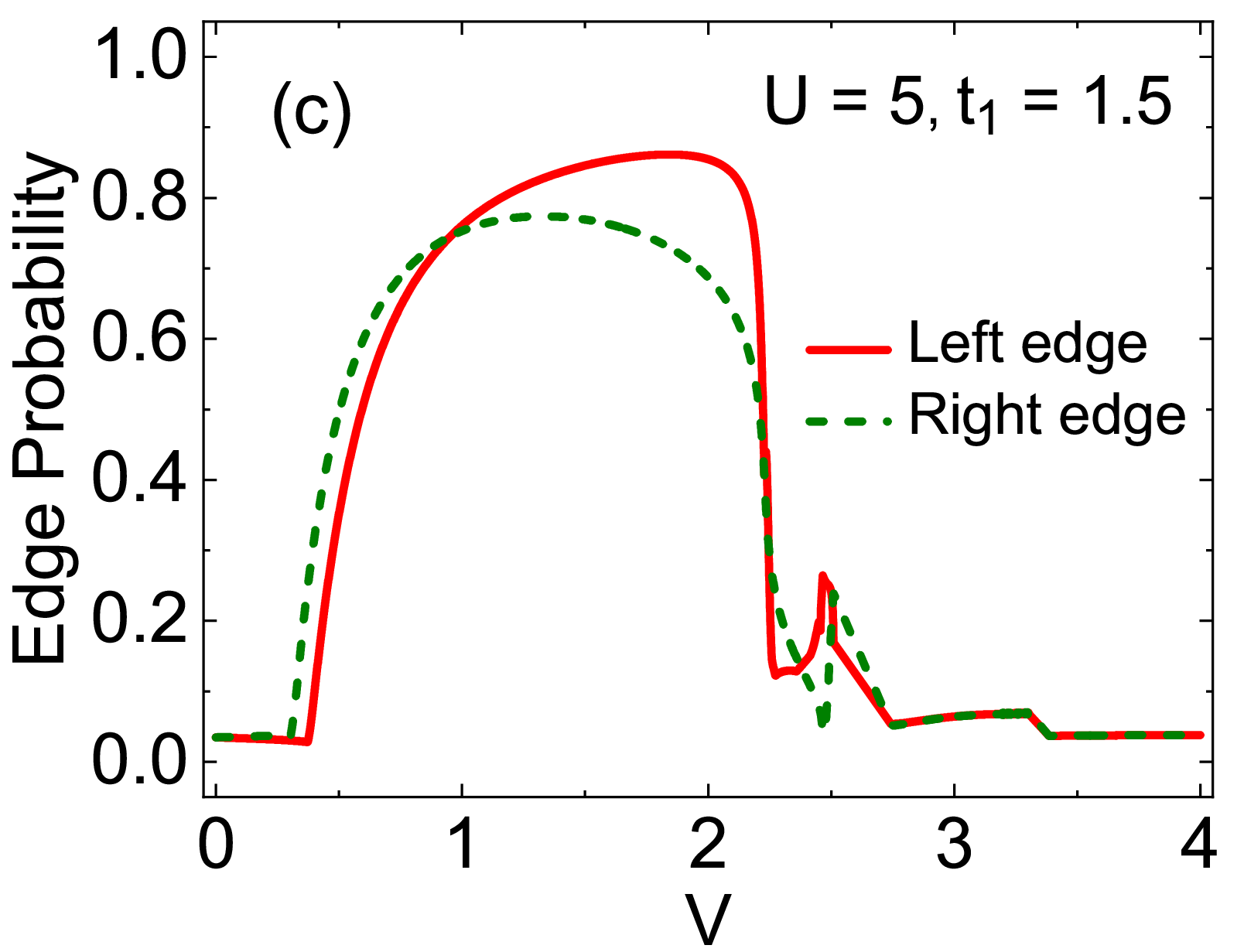}
\caption{(Color online). Maximum left (red) and right (green) edge probability as a function of interaction parameters. (a) Topological phase ($t_1=0.5$), variation with $U$ at fixed $V=0$. (b) Topological phase, variation with $V$ at fixed $U=3$. (c) Trivial phase ($t_1=1.5$), variation with $V$ at fixed $U=5$. the definition of maximum left/right edge probability is mentioned in the text.}
\label{line_space}
\end{figure}

The variation of edge localization across the parameter space is shown in Fig.~\ref{line_space}, where the red and green curves represent the maximum left and right edge probabilities, respectively. For each $(U,V)$ combination, the edge probability of every eigenstate is examined, and the maximum value is taken to characterize the edge localization.

Figure~\ref{line_space}(a) shows the topological phase with $V=0$ as a function of $U$. At $U=V=0$, both curves exhibit high value, confirming the existence of the two SSH edge states. As $U$ increases, the edge probability gradually 
decreases, and the edge modes disappear around $U\approx 1.7$. The red and green curves overlap completely, indicating equal localization weights at the left and right boundaries for all $U$ in this scan.

Figure~\ref{line_space}(b) shows the evolution with $V$ at fixed $U=3$ in the topological phase. At $V=0$, no edge states are present, consistent with the disappearance of the SSH boundary modes above $U\approx 1.7$. Upon increasing $V$, edge localization reemerges at both boundaries, reaches a sharp maximum near $2V\approx U$, and then decays rapidly to zero as $2V$ exceeds $U$. The peak at $2V\approx U$ is significant because it coincides with the boundary between the SDW-dominated ($U>2V$) and CDW-dominated ($U<2V$) regimes, where the competition between the two orders is strongest and edge localization is most pronounced. Once $2V$ exceeds $U$, CDW order dominates and edge states give way to mid-chain domain-wall modes. Here, the two curves do not overlap for most of the $V$ range. Although qualitatively similar, their mismatch indicates unequal localization weights at the left and right edges. This asymmetry arises from the self-consistent on-site potential, which remains symmetric under pure $U$ but develops a left-right imbalance when $V$ is introduced. The resulting unequal potential profiles at the two edges produce the observed mismatch in localization amplitudes.

Figure~\ref{line_space}(c) shows the $V$ scan for the trivial phase ($t_1 = 1.5$) at fixed $U = 5$, with $V$ ranging from $0$ to $4$. The behavior mirrors the topological case, that is, edge probability rises as $V$ increases in the SDW-dominated regime ($U > 2V$), reaches a sharp maximum near $2V\approx U$, and then decays rapidly as $2V$ exceeds $U$ and the system enters the CDW-dominated regime. The peak occurs at a larger $V$ compared to Fig.~\ref{line_space}(b), consistent with the larger $U = 5$ shifting the SDW-CDW boundary to a higher $V$. The red and green curves exhibit the same qualitative behavior as in the topological phase, confirming that edge localization is governed by the ratio $2V/U$ rather than the underlying band topology. We note that no separate $U$-dependent scan is presented for the trivial regime, since pure on-site interaction alone does not produce edge-localized states in this case, as already established. The edge probability scan at the critical point is expected to exhibit the same qualitative behavior and is therefore not shown here.

We note that all localized states reported here vanish under periodic boundary conditions, confirming their origin as boundary or domain-wall bound states rather than bulk localization.

{\it Conclusion}. We have systematically studied the extended Hubbard model on a dimerized SSH chain at half-filling using a self-consistent mean-field approach, examining the interplay of on-site and nearest-neighbor interactions across topological, critical, and trivial hopping regimes. Our main findings are as follows:

$\bullet$ Localized states are governed by the ratio $2V/U$, not by SSH band topology.

$\bullet$ $U>2V$ yields edge SDW modes and $U<2V$ produces mid-chain CDW domain-wall states, universally across all three hopping regimes.

$\bullet$ In the topological phase, a spin-split edge configuration emerges with opposite spins at opposite ends.

$\bullet$ Edge probability peaks sharply at $2V\approx U$ and decays rapidly away from the phase boundary.

These results establish that the localized states are intrinsic correlation-driven excitations of the one-dimensional extended Hubbard model with open boundaries, largely independent of the underlying band topology.


\begin{thebibliography}{99}
\bibitem{ssh1} W. P. Su, J. R. Schrieffer, and A. J. Heeger, {\it Solitons in Polyacetylene}, Phys. Rev. Lett. \textbf{42}, 1698 (1979).

\bibitem{ssh2} W. P. Su, J. R. Schrieffer, and A. J. Heeger, {\it Soliton excitations in polyacetylene}, Phys. Rev. B, \textbf{22}, 4 (1980).

\bibitem{hasan} M. Z. Hasan and C. L. Kane, {\it Colloquium: Topological insulators}, Rev. Mod. Phys. {\bf 82}, 3045 (2010)

\bibitem{jkasb} J. K. Asb\'{o}th, L. Oroszl\'{a}ny, and A. P\'{a}lyi, {\it A Short Course on Topological Insulators}, Volume 919. Springer, Berlin (2016). 

\bibitem{ryu} S. Ryu and Y. Hatsugai, {\it Topological origin of zero-energy edge states in particle-hole symmetric systems}, Phys. Rev. Lett. {\bf 89}, 077002 (2002).

\bibitem{heeger} A. J. Heeger, {\it Nobel Lecture: Semiconducting and metallic polymers}, Rev. Mod. Phys. {\bf 73}, 681 (2001).

\bibitem{lu} L. Lu, J. D. Joannopoulos, and M. Solja\v{c}i\'{c}, {\it Topological photonics}, Nat. Photon. 8, 821 (2014).

\bibitem{ozawa} T. Ozawa, H. M. Price, A. Amo, N. Goldman, M. Hafezi, L. Lu, M. C. Rechtsman, D. Schuster, J. Simon et al., {\it Topological photonics}, Rev. Mod. Phys. {\bf 91}, 015006 (2019).

\bibitem{atala} M. Atala, M. Aidelsburger, J. T. Barreiro, D. Abanin, T. Kitagawa, E. Demler, and I. Bloch, {\it Direct measurement of the Zak phase in topological Bloch bands}, Nat. Phys. {\bf 9}, 795 (2013).

\bibitem{meier} E. J. Meier, F. Alex An, and B. Gadway, {\it Observation of the topological soliton state in the Su-Schrieffer-Heeger model}, Nat. Commun. {\bf 7}, 13986 (2016).

\bibitem{huber} S. D. Huber, {\it Topological mechanics}, Nat. Phys. {\bf 12}, 621 (2016). 

\bibitem{gma} G. Ma, M. Xiao, and C. T. Chan, {\it Topological phases in acoustic and mechanical systems}, Nat. Rev. Phys. {\bf 1}, 281 (2019).

\bibitem{bt} B.-T. Ye, L.-Z. Mu, and H. Fan, {\it Entanglement spectrum of Su-Schrieffer-Heeger-Hubbard model}, Phys. Rev.
B {\bf 94}, 165167 (2016).

\bibitem{barbiero} L. Barbiero, L. Santos, and N. Goldman, {\it Quenched dynamics and spin-charge separation in an interacting topological lattice}, Phys. Rev. B {\bf 97}, 201115(R) (2018).

\bibitem{nhle} N. H. Le, A. J. Fisher, N. J. Curson, and E. Ginossar, {\it Topological phases of a dimerized Fermi-Hubbard model for semiconductor nano-lattices}, npj Quantum Inf. {\bf 6}, 24 (2020).

\bibitem{mikhail} D. Mikhail and S. Rachel, {\it Su-Schrieffer-Heeger-Hubbard model at quarter filling: Effects of magnetic field and nonlocal interactions}, Phys. Rev. B {\bf 110}, 205106 (2024).

\bibitem{yxwang} Y.-X. Wang and Y. Zhong, {\it Ground-state phase diagram of the quarter-filled interacting Su-Schrieffer-Heeger chain}, Phys. Rev. B {\bf 112}, 245122 (2025).


\bibitem{kumar} J. Bisht, S. Jalal, and B. Kumar, Phys. Rev. B, \textbf{110}, 245110 (2024).

\bibitem{tjin} T. Jin, P. Ruggiero, and T. Giamarchi, {\it Bosonization of the interacting Su-Schrieffer-Heeger model}, Phys. Rev. B, \textbf{107}, L201111 (2023).


\bibitem{hubbard} J. Hubbard, {\it Electron Correlations in Narrow Energy Bands}, Proc. Roy. Soc. A, \textbf{276}, 238 (1963).

\bibitem{hirsch} J. E. Hirsch, {\it Charge-Density-Wave Transition in the Extended Hubbard Model}, Phys. Rev. Lett. {\bf 53}, 2327 (1984).

\bibitem{pgj} P. G. J. van Dongen, {\it Extended Hubbard model at weak coupling}, Phys. Rev. B {\bf 50}, 14016 (1994).

\bibitem{brendel} G. Bed\"{u}rftig, B. Brendel, H. Frahm, and R. M. Noack, {\it Friedel oscillations in the open Hubbard chain}, Phys. Rev. B {\bf 58}, 10225 (1998).

\bibitem{tomita} N. Tomita and H. Fukutome, {\it Elementary Defects, Halfons, Connecting the SDW and CDW in the One-Dimensional Extended Hubbard Model}, J. Phys. Soc. Jpn. {\bf 62}, 1634 (1993).

\bibitem{matsuno} G. Matsuno, Y. Omori, T. Eguchi, and A. Kobayashi, {\it Topological Domain Wall and Valley Hall Effect in Charge Ordered Phase of Molecular Dirac Fermion System $\alpha$-(BEDT-TTF)$_2$I$_3$}, J. Phys. Soc. Jpn. {\bf 85}, 094710 (2016).

\bibitem{ohki} D. Ohki, G. Matsuno, Y. Omori, and A. Kobayashi, {\it Domain Wall Conductivity with strong Coulomb interaction of two-dimensional massive Dirac Electrons in the Organic Conductor $\alpha$-(BEDT-TTF)$_2$I$_3$}, J. Phys. Soc. Jpn. {\bf 87}, 054703 (2018).
\end{thebibliography}
\end{document}